\documentclass[prd,aps,showpacs]{revtex4}

\pdfoutput=1

\usepackage{amssymb,amsmath}
\usepackage{epsfig}
\usepackage{color}

\usepackage{hyperref}
\def\hhref#1{\href{http://arxiv.org/abs/#1}{arXiv:#1}} 

\newcommand{\comment}[1]{} 
\newcommand{\nn}{\nonumber}
\def\ll{\mathcal L}
\comment{
}

\begin{document}

\title{The Derivative Expansion at Small Mass for the Spinor Effective Action}


\author{Gerald~V.~Dunne$^{1}$, Adolfo~Huet$^{1}$, Jin Hur$^{2}$ and Hyunsoo Min$^{3}$}
\affiliation{$^{1}$ Department of Physics, University of Connecticut, Storrs, CT 06269-3046, USA
\\
$^{2}$
School of Computational Sciences, Korea Institute for Advanced Study, Seoul 130-012, Korea
\\
$^{3}$
Department of Physics, University of Seoul, Seoul 130-743, Korea}

\begin{abstract}
We study the small mass limit of the one-loop spinor effective action, comparing the derivative expansion approximation with exact numerical results that are obtained from an extension to spinor theories of the partial-wave-cutoff method. In this approach one can compute numerically the renormalized one-loop effective action, for radially separable gauge field background fields in spinor QED. We highlight  an important difference between the small mass limit of the derivative expansion for spinor and scalar theories.
\end{abstract}


\maketitle

\section{Introduction}
The  one-loop effective action in a background field is an important quantity  in quantum field theory \cite{Jackiw:1974cv,Iliopoulos:1974ur}. In gauge theories, the  one-loop effective action has been calculated exactly and analytically only for certain very special gauge field backgrounds, such as those with constant (or, in non-Abelian theories, covariantly constant) field strength, based on the seminal work of Heisenberg and Euler \cite{Heisenberg:1935qt}, and Schwinger \cite{Schwinger:1951nm}; and  for some very special one-dimensional cases where  the field is inhomogeneous \cite{Dunne:2004nc}. The goal of this paper is to extend further the class of background fields for which we can compute the renormalized one-loop effective action. We are particularly interested in studying the small fermion mass limit, motivated by its importance for chiral physics, and also by the fact that while the large mass limit is well understood \cite{Novikov:1983gd}, much less is known about the small mass limit for general gauge backgrounds.

Recently, a new approach has been developed for computing exactly, but numerically, the gauge theory effective action in a class of background fields that permit a separation of variables reducing to a set of one-dimensional operators. This has been explored in most detail for radially separable backgrounds, where the method has been called the "partial-wave cutoff method"
\cite{Dunne:2004sx,Dunne:2005te,Dunne:2006ac,Dunne:2007mt,Hur:2008yg}.  The first application of this method was to the full mass dependence of the quark determinant in an instanton background, yielding a smooth interpolation between the known large and small mass extreme limits \cite{Dunne:2004sx}. Subsequent papers have concentrated on scalar theories. Indeed,  even 
the instanton background computation made use of a special property of self-dual backgrounds that implies that the spinor effective action can be expressed directly in terms of the scalar effective action \cite{'tHooft:1976fv,Jackiw:1977pu,Brown:1977bj,Carlitz:1978xu}. Here, in this paper, we present the spinor approach without assuming self-duality of the background gauge field.

The basic idea of the "partial-wave cutoff method" is simple: the one-loop effective action requires the logarithm of the determinant of an operator, and there is a simple method [known as the Gel'fand-Yaglom method \cite{Gelfand:1959nq}] for computing the determinant of an {\it ordinary} differential operator, without computing its eigenvalues. For a {\it partial} differential operator, if the problem is separable down to a set of {\it ordinary} differential operators, we can formally sum over the Gel'fand-Yaglom results for each term in the separation  sum. The technical difficulty is that this sum is naively divergent, and this must be addressed by a suitable regularization and renormalization. This has been resolved in previous publications for gauge theories with scalars and for self-interacting scalar theories. Here we show how this works for spinor theories with background gauge fields that are radially separable, and not necessarily self-dual.

While this class of radially separable backgrounds is large, and includes important special cases such as instantons, monopoles and vortices, there are of course many other physically important background fields  that are not separable in this way. In these cases we must resort to approximation methods  in order to compute the effective action.  
In order to investigate the region of validity of these approximations, we compare our exact numerical results with two such approximations, the large mass expansion and the derivative expansion, and show that for spinor theories a new feature arises when evaluating the derivative expansion approximation for light fermions.

\section{Partial-wave decomposition of the Dirac operator}


We begin with a chiral decomposition of the effective action. The Euclidean one-loop effective action in spinor QED is the logarithm of the determinant of the corresponding Dirac operator:
\begin{eqnarray}
\Gamma[A] = -\ln \, \det\left( \displaystyle{\not} D+m \right)
=-\frac{1}{2}\ln\,\det\left(-\displaystyle{\not} D^2 +m^2\right)
\label{oneloop}
\end{eqnarray}
Here $\displaystyle{\not} D
= \gamma_\mu( \partial_\mu + i A_\mu (x) )
$ is the Dirac operator in Euclidean 4-dimensional spacetime, and $A_\mu(x)$ is the classical background gauge field. We will use the following standard representation for the $4\times 4$ Dirac matrices in Euclidean spacetime \cite{Jackiw:1977pu}:
\begin{equation}
\gamma_\mu =  \begin{pmatrix}
& 0 & \alpha_\mu  & \\
& \bar{\alpha}_\mu & 0 & \end{pmatrix}
\end{equation}
 where
\begin{eqnarray}
\alpha_\mu =(-i \vec{\sigma}, 1) \quad , \quad  \bar{\alpha}_\mu =
(\alpha_\mu)^\dagger = (i \vec{\sigma}, 1)
\end{eqnarray}
and the $\sigma_i$ are the $2\times 2$ Pauli matrices. Using this Dirac algebra representation we see clearly the chiral decomposition:
\begin{eqnarray}
-\displaystyle{\not} D^2 +m^2=
\begin{pmatrix}m^2+D D^\dagger &0\cr
0& m^2+D^\dagger D
\end{pmatrix}
\end{eqnarray}
where we have defined
\begin{eqnarray}
D\equiv \alpha_\mu D_\mu\qquad , \qquad D^\dagger \equiv - \bar\alpha_\mu D_\mu
\label{d}
\end{eqnarray}
Thus, we write
\begin{eqnarray}
\Gamma[A] &=&-\frac{1}{2}\ln\,\det\left( m^2+D D^\dagger \right) -\frac{1}{2}\ln\,\det\left( m^2+D^\dagger D  \right)\nonumber\\
&\equiv& \Gamma^{(+)}[A]+\Gamma^{(-)}[A]
\label{sum}
\end{eqnarray}
For later use, we recall the familiar properties of the $\alpha_\mu$ matrices:
\begin{eqnarray}
m^2+D D^\dagger &=& m^2-D_\mu^2+\frac{1}{2}F_{\mu\nu} \bar{\eta}^a_{\mu\nu}\,\sigma_a
\label{dddag}\\
m^2+D^\dagger D &=& m^2-D_\mu^2+\frac{1}{2} F_{\mu\nu} \eta^a_{\mu\nu}\,\sigma_a
\label{ddagd}
\end{eqnarray}
where $\eta^a_{\mu\nu}$ and $\bar\eta^a_{\mu\nu}$ are the 't Hooft tensors \cite{'tHooft:1976fv,Jackiw:1977pu}. 

Now we make the following simple observation, that we can write the renormalized effective action as
\begin{eqnarray}
\Gamma_{\rm ren}[A] &=& 2\Gamma^{(\pm)}_{\rm ren}[A]
\mp\left(\Gamma^{(+)}_{\rm ren}[A]-\Gamma^{(-)}_{\rm ren}[A]\right)
\label{choice}
\end{eqnarray}
Furthermore, we know that the difference of the renormalized effective action for the two chiralities takes a special form, related to the chiral anomaly:
\begin{eqnarray}
\Delta\Gamma_{\rm ren}[A] &\equiv& \left(\Gamma^{(+)}_{\rm ren}[A]-\Gamma^{(-)}_{\rm ren}[A]\right)\nonumber\\
&=&\frac{1}{2}\frac{1}{(4\pi)^2}\ln \left(\frac{m^2}{\mu^2}\right)\int d^4 x\, F_{\mu\nu} F^*_{\mu\nu}
\label{difference}
\end{eqnarray}
Thus, to evaluate the spinor effective action it is sufficient to evaluate the effective action for just one of the chiralities. That is, we can compute either $\Gamma^{(+)}$ or $\Gamma^{(-)}$, but we do not need to compute both \cite{Hur:2010bd}.

This is a useful observation because there exist background fields for which the computation is  significantly easier for one chirality than for the other. For example, if the background field is self-dual then $F_{\mu\nu} \bar{\eta}^a_{\mu\nu}=0$, since $\bar{\eta}^a_{\mu\nu}$ is anti-self-dual. Therefore the positive chirality operator reduces to the scalar Klein-Gordon operator:
\begin{eqnarray}
m^2+D D^\dagger &=& m^2-D_\mu^2
\label{kg}
\end{eqnarray}
which implies that
\begin{eqnarray}
\Gamma_{\rm spinor}[A]=-2 \Gamma_{\rm scalar}[A]-\Delta\Gamma[A]
\label{spsc}
\end{eqnarray}
This familiar result enables the computation of the quark determinant in an instanton background via the associated scalar determinant, which has a partial-wave decomposition \cite{'tHooft:1976fv,Brown:1977bj,Carlitz:1978xu,Dunne:2004sx,Dunne:2005te}.

In this paper we study a special class of background gauge fields that are not self-dual, but for which the Dirac operator still admits a partial wave decomposition \cite{Adler:1972qq,Adler:1974nd,Bogomolny:1981qv,Bogomolny:1982ea,Fry:2003uy,Fry:2010cd}. Before coming to the radial decomposition, we first note that these gauge fields admit a simple chiral decomposition.
Specifically, we consider gauge fields of the form
\begin{eqnarray} \label{Apotential}
A_\mu (x) &=& \eta_{\mu \nu}^3 x_\nu g(r) \quad ,
\label{a}
\end{eqnarray}
where $\eta_{\mu \nu}^3$ are the 't Hooft symbols \cite{'tHooft:1976fv}, and $g(r)$ is a radial profile function, to be specified below.
This type of background field is symmetric under $O(2) \times O(3)$ 
transformations,
leading to a partial wave decomposition
\cite{Bogomolny:1981qv,Bogomolny:1982ea}. This decomposition applies for any radial profile function $g(r)$, so we have the freedom to study a wide class of background fields. In particular, we can investigate the role of zero modes, the presence or absence of which depends on the form of $g(r)$.
The associated field strength tensor is
\begin{equation}
F_{\mu \nu}(x) = - 2 \eta_{\mu \nu}^3 g(r) - \frac{g'(r)}{r} \left(\eta_{\mu \sigma}^3 x_\nu x_\sigma   -  \eta_{\nu \sigma}^3 x_\mu x_\sigma\right)
\label{f}
\end{equation}
Note that this field strength is not self-dual, due to the presence of the term proportional to $g'(r)$. Thus, the positive chirality operator $DD^\dagger$ does not simplify as in (\ref{kg}). However, for this field the negative chirality operator does take a particularly simple form:
\begin{eqnarray}
m^2+D^\dagger D=m^2-D_\mu^2+\left(4g(r)+r\,g'(r)\right)\sigma_3
\label{negative}
\end{eqnarray}
This is diagonal in spinor degrees of freedom, so if we work in the negative chirality sector, then we can immediately use our previous results for scalar Klein-Gordon operators \cite{Dunne:2004sx,Dunne:2005te,Dunne:2006ac,Dunne:2007mt,Hur:2008yg}, just by including an extra "potential" equal to $\pm \left(4g(r)+r\,g'(r)\right)$. Thus, we can write
\begin{eqnarray} \label{radial1} 
m^2+D^\dagger D&=&-\left[ \partial_r^2  + \frac{4 l + 3}{r} \partial_r  - r^2 g(r)^2 - 4 g(r) l_3   - m^2
\mp (4 g(r) + r g'(r) ) \right] 
\nonumber\\
&\equiv&
m^2+\mathcal{H}_{(l, l_3, s)}
\end{eqnarray}
where the quantum number $l$ takes half-integer values: $l=0, \frac{1}{2}, 1, \frac{3}{2}, \dots$, while $l_3$ ranges from $-l$ to $l$, in integer steps. In this way, we  compute $\Gamma^{(-)}[A]$, and then the full spinor effective action can then be obtained using (\ref{choice}) and (\ref{difference}).

\section{The partial-wave cutoff method}

In this section we briefly recall the partial-wave cutoff method developed previously for radially separable fields of the form (\ref{a}), adapted to the negative chirality sector of the spinor theory using (\ref{negative}). The basic idea of the partial-wave cutoff method involves separating the sum over the the quantum number $l$ into a low partial-wave contribution, each term of which is computed using the (numerical) Gelfand-Yaglom method, and a high partial-wave contribution, whose sum is computed analytically using a WKB expansion. The regularization and renormalization procedure tells us how to combine these two contributions to yield the finite and renormalized effective action \cite{Dunne:2004sx,Dunne:2005te,Dunne:2006ac,Dunne:2007mt,Hur:2008yg}.

\subsection{Low partial-wave contribution}
The low partial-wave contribution for our system is given by 
\begin{eqnarray}
\Gamma^{(-)}_{{\rm{L}}} &=& -\sum_{s = \pm} \; \sum_{l =
0, \frac{1}{2} , 1, \ldots}^L \; \Omega(l) \sum_{l_3 = -l}^{l}
\ln \left( \frac{ {\rm{det}}(m^2+\mathcal{H}_{(l, l_3,
s)})}{{\rm{det}}(m^2+\mathcal{H}_{(l, l_3, s)}^{{\rm{free}}})} \right) \, ,
\end{eqnarray}
where $L$ is an arbitrary angular momentum cutoff.
The factor $
\Omega(l) =  (2 l +1) $
is  the degeneracy factor, and the
$s$ sum comes from adding the contributions of each spinor
component in  \eqref{radial1}.  In order to evaluate this quantity, we
use the Gel'fand-Yaglom method, which we now briefly review (for further details see \cite{Dunne:2004sx,Dunne:2005te,Dunne:2006ac}). Let
$\mathcal{M}_1$ and $\mathcal{M}_2$ denote two second-order radial
differential operators on the interval $r \: \in \, [\, 0,\infty)$
and let $\Phi_1(r)$ and $\Phi_2(r)$ be solutions to the following
initial value problem :
\begin{equation}
\mathcal{M}_i \Phi_i(r) = 0; \quad \Phi_i(r) \sim r^{2 l} \quad {\rm{as}} \quad r \to 0 \, .
\label{initial}
\end{equation}
Then the ratio of the determinants is given by
\begin{eqnarray}
\frac{{\rm{det}} \mathcal{M}_1}{{\rm{det}} \mathcal{M}_2} &=& \lim_{ R \to \infty} \left(  \frac{\Phi_1(R)}{\Phi_2(R)}      \right) \, .
\end{eqnarray}
In the negative chirality sector we can take, $\mathcal{M}_1 =m^2+\mathcal{H}_{(l, l_3, s)}$, and
$\mathcal{M}_2$ to be the corresponding free operator: i.e, the same operator with the background field set to zero: $g(r)\equiv 0$.
Thus, for a given radial profile function $g(r)$ in (\ref{a}), for each value of $(l, l_3, s)$ we need to solve 
\begin{eqnarray}
\Phi_{\pm}''(r) + \frac{4 l + 3 }{r} \Phi_{\pm}'(r) - \left( m^2 +
4 l_3 g(r) + r^2 g(r)^2 \mp [4 g(r) + r g'(r)]   \right)
\Phi_{\pm}(r) &=& 0 \, ,
\end{eqnarray}
with the initial value boundary condition in (\ref{initial}). The value of $\Phi$ at $r=\infty$ gives us the value of the determinant for that partial wave. In fact, 
the corresponding free equation is analytically soluble.
It is numerically more convenient to define
\begin{eqnarray}
S_\pm^{(l, l_3 )}(r) &\equiv& \ln \left( \frac{\Phi_{l, l_3,\pm} (r)}{\Phi_{l, l_3,\pm}^{\rm free} (r)} \right) \, ,
\end{eqnarray}
and solve numerically the corresponding initial value problem for $S(r)$, as explained in \cite{Dunne:2005te,Dunne:2007mt}.
Then the contribution of the low-angular-momentum partial-waves to the effective action is
\begin{eqnarray}
\Gamma^{(-)}_{{\rm{L}}} &=& - \sum_{l = 0, \frac{1}{2}, 1 ,
\ldots}^L \; \Omega(l) \sum_{l_3 = -l}^{l} [ S_{+}^{(l,
l_3)}(\infty) + S_{-}^{(l, l_3)}(\infty) ] \, .
\end{eqnarray}
While each term in the sum over $l$ is finite and simple to compute, the sum over $l$  is divergent as $L \to \infty$. In fact,  only after
 adding the renormalized  contribution of the high partial-wave modes  do we obtain a finite and renormalized result for the effective action.

\subsection{High partial-wave contribution}
The high partial-wave contribution for our system is given by
\begin{eqnarray} \label{gammalow}
\Gamma^{(-)}_{{\rm{H}}} &=& -\sum_{s = \pm \frac{1}{2}} \; \sum_{l = L
+ \frac{1}{2}}^\infty \; \Omega(l) \sum_{l_3 = -l}^{l} \ln
\left( \frac{ {\rm{det}}(m^2+\mathcal{H}_{(l, l_3, s)})}{{\rm{det}}(m^2+\mathcal{H}_{(l, l_3, s)}^{{\rm{free}}} )}
\right) \, .
\end{eqnarray}
Again, since by (\ref{negative}) we only need the negative chirality sector, and the negative chirality sector diagonalizes fully, we can apply our previous scalar analysis to each of the diagonal components, adding the appropriate term $\pm (4g+rg')$ to the Klein-Gordon operator. This modifies the detailed expressions as follows. In the large $L$ limit, we have
\begin{align}
\Gamma^{(-)}_{{\rm{H}}} &= \int_0^\infty dr \left( \frac{8 \, g(r) r^3}{3 \sqrt{\tilde{r}^2 + 4}} \right) L^2
+ \int_0^\infty dr \left( \frac{2 r^3 ( 3 \tilde{r}^3 + 8) g(r)^2}{( \tilde{r}^2 + 4)^{3/2}} \right) L \\
&+ \int_0^\infty dr \Bigg\{ \frac{r^3}{45 ( \tilde{r}^2 +
4)^{7/2}} \Bigg[ -6 r^4 (5 \tilde{r}^4 + 28 \tilde{r}^2 + 32 )g(r)^4
\\ &+ 15 (33 \tilde{r}^6 +  335 \tilde{r}^4  + 1192 \tilde{r}^2 + 1600) g(r)^2 \\
&+ 10 r (15 \tilde{r}^6 + 184 \tilde{r}^4 + 776 \tilde{r}^2 + 1120) g(r)g'(r) \\
&+ 5 r^2 (3 \tilde{r}^6 + 38 \tilde{r}^4+ 160 \tilde{r}^2 + 224) g'(r)^2 + 20 r^2 (4 + \tilde{r}^2)^2
g(r)g''(r)]
 \Bigg] \\
&+ \frac{r^3 (20 g(r)^2 + 10 g(r)g'(r)r + g'(r)^2 r^2 ) }{12} \Bigg[
\gamma + 2  \ln L  - 2 \ln \Bigg( \frac{r}{2 + \sqrt{\tilde{r}^2 +
4}} \Bigg)
 \Bigg]
  \Bigg\} \\
&-  \frac{r^3 (20 g(r)^2 + 10 g(r)g'(r) r + g'(r)^2 r^2 )}{12 \,
\epsilon} + O\Bigg( \frac{1}{L} \Bigg) \, ,
\end{align}
where $\tilde{r} \equiv  \frac{r m}{L}$. We now identify the
counterterm as
\begin{equation}
\delta \Gamma^{(-)} = \frac{1}{12} \Bigg( \frac{1}{\epsilon} - \gamma - 2
\ln \mu  \Bigg) \int_0^\infty dr \; r^3 (20 g(r)^2 + 10 g(r)g'(r) r +
g'(r)^2 r^2 ) \, ,
\label{ct}
\end{equation}
Where $\mu$ is the renormalization scale.
Note that
\begin{eqnarray}\label{FF}
\frac{1}{2} \int_0^\infty dr \; r^3 (F_{\mu \nu} F_{\mu \nu})  &=&
\int_0^\infty dr \; r^3 (8 g(r)^2 + 4 g(r)g'(r) r +
g'(r)^2 r^2 ) \, ,  \\
\frac{1}{2} \int_0^\infty dr \; r^3 (F_{\mu \nu} \tilde{F}_{\mu \nu})  &=&
\int_0^\infty dr \; r^3 (8 g(r)^2 + 4 g(r)g'(r) r  ) \, ,
\label{FFdual}
\end{eqnarray}
and thus, the counterterm corresponds to the following combination:
\begin{equation}
\delta \Gamma^{(-)} = \frac{1}{24} \Bigg( \frac{1}{\epsilon} - \gamma - 2
\ln \mu  \Bigg)  \int_0^\infty dr \; r^3 \Bigg( F_{\mu \nu} F_{\mu \nu} + \frac{3}{2} F_{\mu \nu} \tilde{F}_{\mu \nu}  \Bigg) \, .
\end{equation}
The appearance of the $F_{\mu \nu} \tilde{F}_{\mu \nu}$ term
is a special feature of the spinor calculation that does not occur in the scalar case.

Having identified the counter-term, we can now write an explicit expression for the large $L$ behavior of the high partial-wave contribution to the renormalized effective action:
\begin{eqnarray}
\Gamma_{{\rm{H}}}^{(-)} = \int_0^\infty dr  \left( Q_{{\rm {log}}}(r)\ln L + \sum_{n = 0}^2 Q_n ( r ) L^n  +  \sum_{n = 1}^N Q_{-n}( r ) \frac{1 }{ L^n }  \right) + O(\frac{1}{L^{n+1}}) \, ,
\label{gammaren}
\end{eqnarray}
with the following expansion coefficients:
\begin{align}
Q_2(r) &=  \frac{8 \, g(r) r^3}{3 \sqrt{\tilde{r}^2 + 4}} \nn \\
Q_1(r) &= \frac{2 r^3 (3 \tilde{r}^3 + 8) g(r)^2}{(
\tilde{r}^2 + 4)^{3/2}} \nn \\
Q_{{\rm {log}}}(r) &= -\frac{1}{6}\, r^3 (20 g(r)^2 + 10 g(r)g'(r) r + g'(r)^2
r^2 ) \nn \\
Q_0(r) &= \frac{r^3}{45 ( \tilde{r}^2 +
4)^{7/2}} \Bigg[ -6 r^4 (5 \tilde{r}^4 + 28 \tilde{r}^2 + 32 )g(r)^4
\nn \\ &+ 15 (33 \tilde{r}^6 +  335 \tilde{r}^4  + 1192 \tilde{r}^2 + 1600) g(r)^2  \nn \\
&+ 10 r (15 \tilde{r}^6 + 184 \tilde{r}^4 + 776 \tilde{r}^2 + 1120) g(r)g'(r) \nn \\
&+  5 r^2 (3 \tilde{r}^6 + 38 \tilde{r}^4+ 160 \tilde{r}^2 + 224) g'(r)^2 + 20 r^2 (4 + \tilde{r}^2)^2
g(r)g''(r)  \Bigg]
 \nn \\
&- Q_{{\rm {log}}}(r)   \ln \Bigg( \frac{ \mu r}{2 + \sqrt{\tilde{r}^2 + 4}}
\Bigg)
 \nn \\
Q_{(-1)}(r) &= -\frac{r^3}{4 ( \tilde{r}^2 + 4)^{9/2}} \Bigg[ 6
r^4( \tilde{r}^6 + 4 \tilde{r}^4 )g(r)^4 \nn\\
&+2 r^2(  \tilde{r}^6 + 16 \tilde{r}^4 + 80\tilde{r}^2 +
 128)g'(r)^2 \nn  \\
&+ (-4 \tilde{r}^8 + 89 \tilde{r}^6 + 1104 \tilde{r}^4 + + 3456 \tilde{r}^2 + 5120 ) g(r)^2 \nn \\
&+16 r( 2 \tilde{r}^6 + 21 \tilde{r}^4 + 92
\tilde{r}^2 + 160)g(r) g'(r) \nn \\
&-4 r^2(\tilde{r}^6 + 8 \tilde{r}^4 + 16
\tilde{r}^2 )g(r) g''(r)
 \Bigg] \, .     \label{spinorQs}
\end{align}
Note that $\Gamma_{{\rm{H}}}^{(-), \rm{ren}}$ involves $L^2$, $L$ and
$\ln L$ terms  that diverge as $L \to \infty$, but these divergences
exactly cancel those of the $\Gamma_{{\rm{L}}}$ contribution (coming from the numerical
Gel'fand-Yaglom computation for the low partial-wave modes),
yielding a finite renormalized result for the effective action. This method is essentially
exact (up to numerical precision) and accurately provides the value of the effective action
for any value of the mass:
\begin{equation}
\Gamma^{(-)}_{\rm{ren}} (m) =  \Gamma_{{\rm{H}}}^{(-), \rm{ren}} (m)+ \Gamma^{(-)}_{{\rm{L}}} (m)    \quad .
\end{equation}

The effective action calculated as above is finite for any non-zero value of the mass, however, from eq.~(\ref{spinorQs}), we note that $Q_0 (r)$ contains a term proportional 
to $\ln \mu$ and therefore when we use "on-shell" renormalization ($\mu = m$), we have
\begin{equation}
\Gamma^{\rm{ren}} (m) \sim \Bigg( - \int_0^\infty dr \, Q_{{\rm {log}}}(r) \Bigg) \ln m  \quad \quad ; \quad \quad   m \to 0 \, .
\end{equation}
In order to analyse the mass dependence of the effective action, we introduce
a modified effective action defined as
\begin{equation}
\tilde{\Gamma}^{\rm{ren}} (m) \equiv   \Gamma^{\rm{ren}} (m) + \Bigg(  \int_0^\infty dr \, Q_{{\rm {log}}}(r) \Bigg) \ln \mu  \,.
\end{equation}
which is independent of the renormaliztion scale $\mu$, and which is
finite at $m=0$.

\section{Results for specific background profiles } \label{results}

Within the class of background gauge fields (\ref{Apotential}), we still have the freedom to specify the radial profile function $g(r)$. Here we note that the large $r$ behavior of $g(r)$ 
determines the presence or absence of zero-modes. To be specific,
the integral of $F_{\mu \nu} \tilde{F}_{\mu \nu}$ counts the
number of zero-modes. From  (\ref{FFdual}) we have:
\begin{eqnarray}
\frac{1}{2} \int_0^\infty dr \; r^3 (F_{\mu \nu} \tilde{F}_{\mu \nu})  &=&
\int_0^\infty dr \; r^3 (8 g(r)^2 + 4 g(r)g'(r) r ) \, , \nn \\
&=& 2 (g(r) r^2)^2 \Big\vert_0^{\infty}  \, .
\end{eqnarray}
Therefore, as long as $g(r)$ falls faster than $1/r^2$ we don't have
zero-modes. In this paper, we study the background fields using two different
 profile functions to set the background field. The first one is
\begin{equation} \label{g1}
g_1 (r) \equiv B (1 - {\rm {Tanh}}( \beta (\sqrt{B} r  - \xi ) )) \, ,
\end{equation} 
where $B$, $\beta$ and $\xi$ are parameters that control the amplitude, range and steepness of the potential. We may call $g_1(r)$ the  "step potential". Since $g_1(r)$
falls off exponentially fast as $r \to \infty$, this case
is free of zero-modes.
\begin{figure}[htb]
\begin{center}
\includegraphics[width=0.6\textwidth]{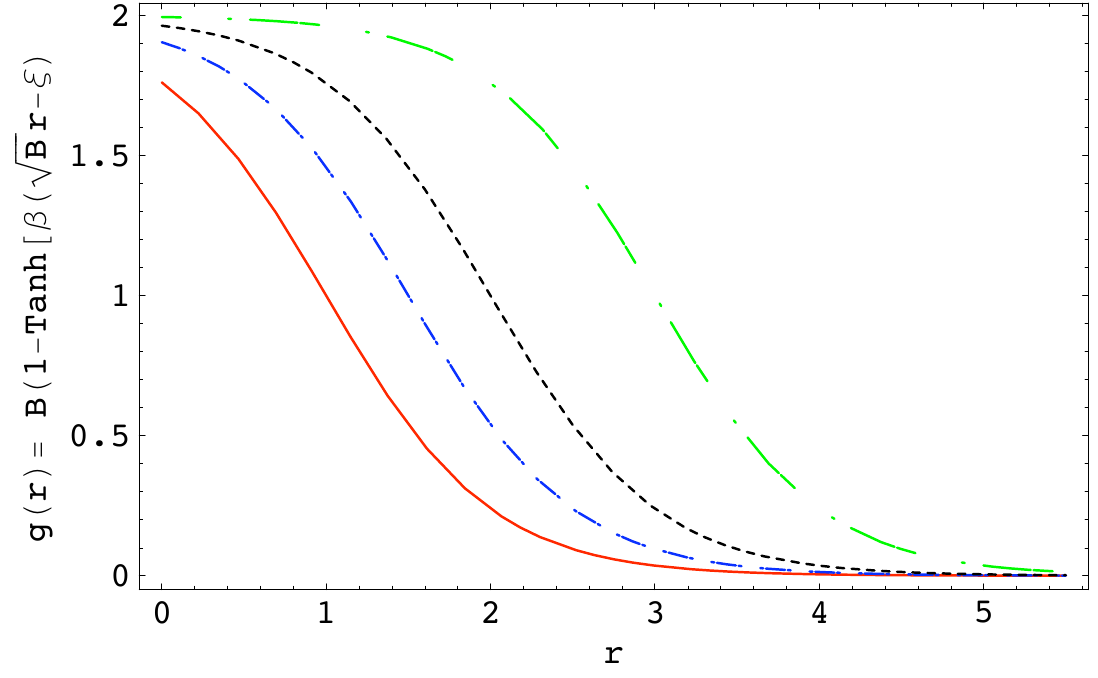}
\includegraphics[width=0.6\textwidth]{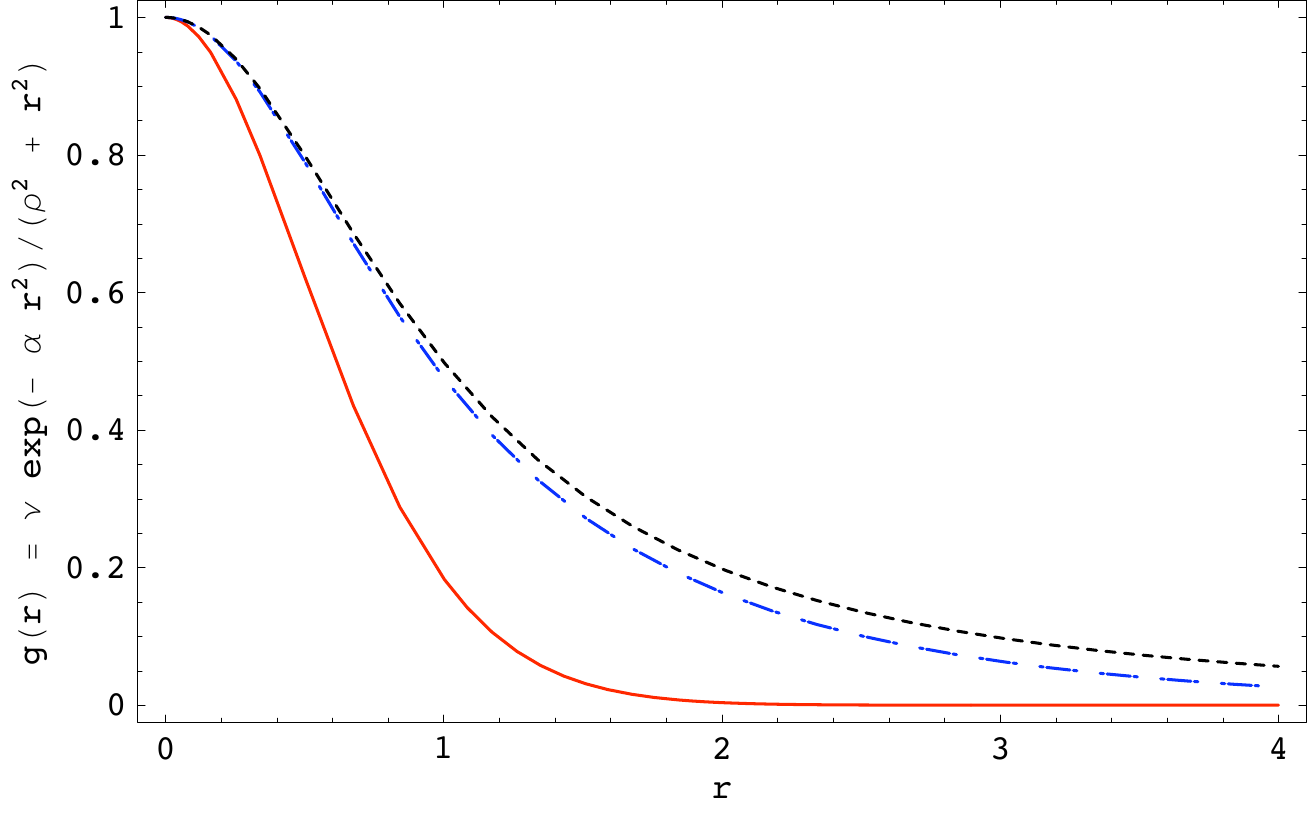}
\end{center}
\caption{ The upper figure shows the radial profile function: $g_1(r) = B (1 - {\rm {Tanh}}( \beta (\sqrt{B} r  - \xi ) )
)$. We have fixed $B=1$ and $\beta=1$ in all the curves. The parameter $\xi$ which controls the \emph{range} 
of the potential is varied. The lowest curve  corresponds to $\xi=1$ (solid/red),
 followed by  curves  corresponding to $\xi=3/2$ (dashes/blue),  $\xi=2$ (dots/black), $\xi=3$ (dot-dashes/green).
 The  lower figure shows a second radial profile function: $g_2 (r) = \nu e^{- \alpha r^2}/(\rho^2 + r^2)$. We have fixed $\nu=1$ and $\rho=1$ in all the curves, the lowest curve  corresponds to $\alpha = 1$ (solid/red),
 followed by  curves  corresponding to $\alpha = 1/20$ (dashes/blue), and $\alpha = 1/400 $ (short-dashes/black). } 
 \label{fig1}
\end{figure}

The second profile function we use is
\begin{equation} \label{g2}
g_2 (r) \equiv \nu\,  e^{-\alpha r^2}/(\rho^2 + r^2) \, , 
\end{equation}
where  $\nu$,
$\rho$ and $\alpha$ are parameters that control the amplitude, range  and steepness of the potential.  This potential has the  following properties:
\begin{eqnarray}
\alpha > 0   &\Longrightarrow& \quad \int d^4 x F_{\mu \nu} F_{\mu \nu} < \infty 
\quad , \quad \int d^4 x F_{\mu \nu} \tilde{F}_{\mu \nu} = 0 \nn \\
\alpha = 0   &\Longrightarrow& \quad \int d^4 x F_{\mu \nu} F_{\mu \nu} \to \infty 
\quad , \quad 0 < \Big| \int d^4 x F_{\mu \nu} \tilde{F}_{\mu \nu} \Big| < \infty  \, . 
\end{eqnarray} 
Note  that $g_2(r)$ goes like $1/r^2$
when we set  $\alpha = 0$, so that in this case there are zero modes. However, whenever the profile function $g(r)$ decays  as $1/r^2$ we have
\begin{eqnarray}
\frac{1}{2} \int_0^\infty dr \; r^3 F_{\mu \nu} F_{\mu \nu}  &=&
\int_0^\infty dr \; r^3 (8 g(r)^2 + 4 g(r)g'(r) r +
g'(r)^2 r^2 ) \, ,\nn  \\
&=&  \int_0^\infty dr \; r^3 F_{\mu \nu} \tilde{F}_{\mu \nu} + \int_0^\infty dr \; r^3 ( g'(r)^2 r^2 ) 
\end{eqnarray} 
which diverges logartihmically.
In Figure \ref{fig1}  we show some plots of our chosen profile functions
$g_1(r)$ and $g_2(r)$ for different values of their parameters.

Our calculational method allows one to compute the effective action for arbitrary values of the fermion mass $m$.
This provides us with a unique opportunity to study the validity of various approximate methods that yield estimates in the limits of large or small mass. In particular, we are able to probe exactly the $m\to 0$ limit, which is of great interest in spinor theories but which is 
notoriously inaccessible by approximate means.
\section{Approximation methods} \label{approximations}

\subsection{The large-mass expansion}
\begin{figure}[htb]
\begin{center}
\includegraphics[width=0.75\textwidth]{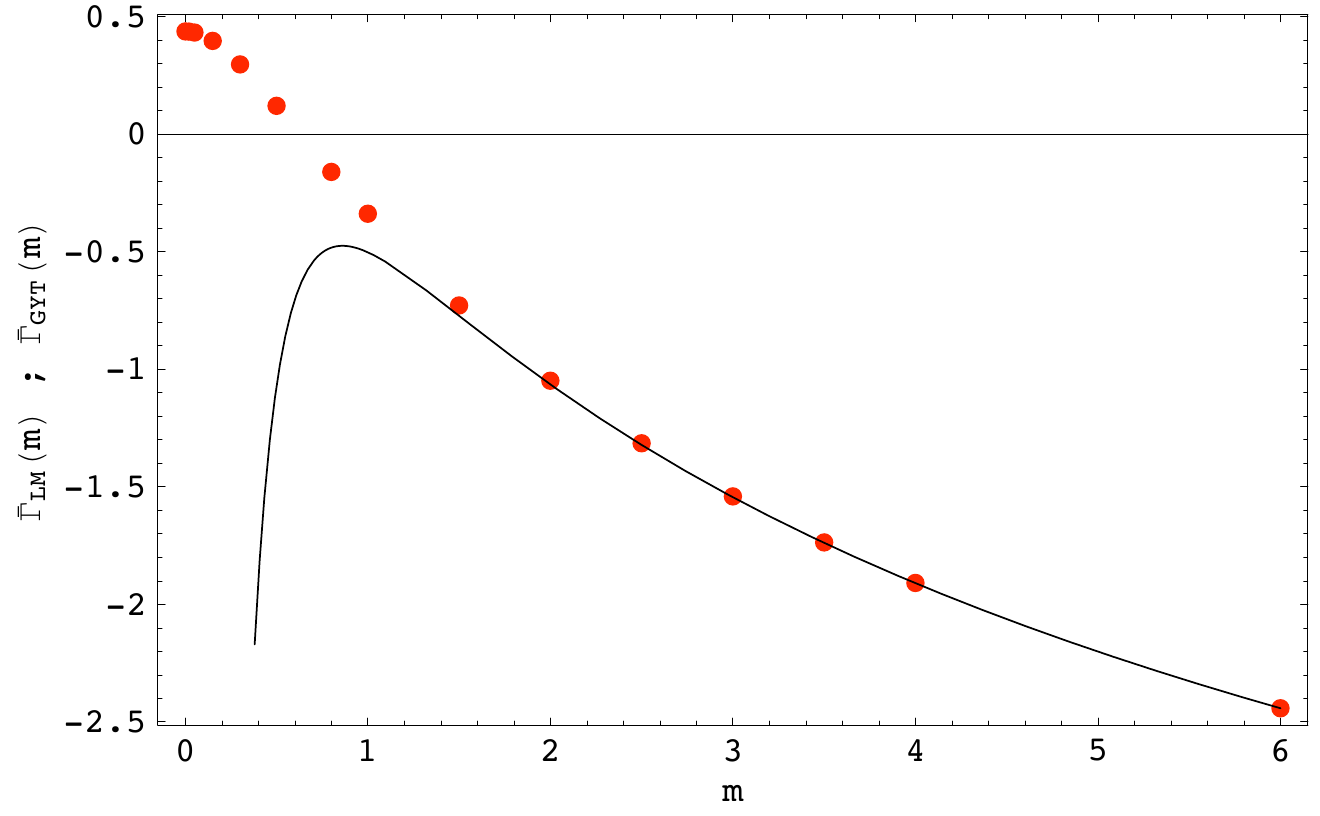}
\end{center}
\caption{ The graph shows  the one-loop effective action, $\tilde{\Gamma} (m)$, as a function of fermion mass $m$,   with the radial profile function:  $g_1(r) = B (1 - {\rm
{Tanh}}( \beta (\sqrt{B} r  - \xi ) ))$.  We have set  $B=1$, $\beta=1$ and $\xi=1$. The 
dots correspond to our exact numerical results, based on the Gelfand-Yaglom theorem,  $\tilde{\Gamma}_{{\rm GYT}}(m)$, and  the solid line shows 
the outcome of the larg-mass expansion $\tilde{\Gamma}_{{\rm LM}}(m)$. }
 \label{fig2}
\end{figure}

The large-mass expansion is the most general approximation method in the
sense that that it may be applied to calculate the effective action
for any well-behaved background. Its main limitation is that it only
applies for large values of the mass and, in fact, it diverges as
$m \to 0$. Thus, it is not directly useful for probing issues related to massless quarks.
To outline this method, consider for instance, the spinor
case
\begin{equation}
\Gamma =
 -\frac{1}{2}\ln {\rm  {det}} ( - {\displaystyle{\not} D}^2 + m^2) \nn \\
 =   \frac{1}{2}\int_0^{\infty} \frac{ds}{s} e^{- m^2 s }\, {\rm {Tr}}\, e^{- s (-{\displaystyle{\not}
 D}^2)} \, .
\end{equation}
In order to analyse the large-mass limit, we may take the small $s$
limit in the proper-time integral and expand the trace factor in
powers of $s$, yielding the heat kernel or Seeley-De Witt expansion:
\begin{equation}
 s \to 0 \quad :  \quad {\rm {Tr}}\, e^{- s (-{\displaystyle{\not}
D}^2)} \sim \frac{1}{(4 \pi s )^{d/2}} \sum_{n = 0}^{\infty}  s^n a_n [
F_{\mu \nu} ] \, .
\end{equation}
The Seeley-DeWitt coefficients $a_n$ are given by Lorentz traces of
powers of $F_{\mu \nu }$and its derivatives. 
For the $A_{\mu} (x)$ backgrounds given by (\ref{g1}) and (\ref{g2})  the Seeley-DeWitt coefficients are proportional to those found in the scalar case:
\begin{eqnarray}
a_1 &=& 0  \, , \nn \\
a_2 &=& \frac{2}{3} \int d^4x\, F_{\mu \nu}  F_{\mu \nu}
\, , \nn \\
a_3 &=& -\frac{ 3}{45} \int d^4x \, (D_\mu F_{\nu
\lambda})(D_\mu F_{\nu \lambda})  \, .
\end{eqnarray}
 Since we are
considering potentials of the form (\ref{Apotential}), with field strength (\ref{f}), we find an expansion of the renormalized effective action as:
\begin{equation}
\tilde{\Gamma}^{{\rm ren}} = \tilde{\Gamma}_{{\rm LM}}^{(0)} \ln m + \tilde{\Gamma}_{{\rm LM}}^{(2)} \frac{1}{m^2} + \tilde{\Gamma}_{{\rm LM}}^{(4)} \frac{1}{m^4} + \cdots
\end{equation} 
The  first two coefficients are:
\begin{eqnarray}
\tilde{\Gamma}_{{\rm LM}}^{(0)} &=& \frac{1}{6} \int_0^\infty dr \; r^3 (8 g(r)^2 + 4 g(r)g'(r) r +
g'(r)^2 r^2 ) \, , \nn  \\
\tilde{\Gamma}_{{\rm LM}}^{(2)} &=& \frac{1}{180} \int_0^\infty dr \big[ 24 r^2 g(r) \big( 15  g'(r) + r \big( 9 g''(r) + r g^{(3)}(r) \big)      \big)   \nn \\
&\, & + r^3 \big( 221 g'(r)^2 + 9 r^2 g''(r)^2 + 2 r g'(r) \big( 71 g''(r) + 6 r g^{(3)}(r) \big)      \big) ] 
\end{eqnarray}
Figure \ref{fig2} presents a comparison between the effective action, as calculated using large-mass expansion and its
exact value obtained using our partial-wave cutoff method. As expected, both methods agree well for large masses. 
\subsection{The derivative expansion}

Another widely used approximation in gauge theories is the derivative expansion. This method is based on the fundamental result that for background gauge fields with constant field strength $F_{\mu\nu}$ it is possible to compute the renormalized effective action in a  simple analytic form \cite{Heisenberg:1935qt,Schwinger:1951nm,Dunne:2004nc}. One can then expand around this soluble case, leading to an expansion of the form \cite{Cangemi:1994by,Gusynin:1995bc,Salcedo:2000hp}
\begin{equation}
S_{{\rm eff}}  \approx  S_0 [F] + S_2 [F, (\partial F)^2] + S_4 [F, (\partial F)^2, (\partial F)^4 ] + \cdots
\label{der}
\end{equation}
The leading term in this expansion is the well-known one-loop effective action for constant backgrounds, first computed by Heisenberg and Euler \cite{Heisenberg:1935qt}.
In euclidean QED, the corresponding one-loop effective {\it Lagrangians} for spinor and scalar QED are
\begin{eqnarray}
\ll_{{\rm spinor}}( a, b)
&=&  -\frac{1}{8 \pi^2} \int_0^{\infty} \, \frac{ds}{s^3}\, e^{-m^2 s }
\Big\{ a b s^2\, {\rm coth}(a s)  {\rm coth}(b s)
+ 1 - \frac{ s^2}{3}\,( a^2 + b^2 ) \Big\} \, , \label{euclidean-EHsp}
\end{eqnarray}
and
\begin{eqnarray}
\ll_{{\rm scalar}}( a, b)
&=&  \frac{1}{16 \pi^2} \int_0^{\infty} \, \frac{ds}{s^3}\, e^{-m^2 s }
\Bigg\{ \, \frac{ a b\, s^2}{{\rm sinh}( a s)  {\rm sinh}( b s)}
- 1 + \frac{ s^2}{6}\,( a^2 + b^2 ) \Bigg\} \label{euclidean-EHsc}  \,,
\end{eqnarray}
where $\pm i a$ and $\pm b$ are the eigenvalues of the $4 \times 4$ constant matrix $F_{\mu \nu}$, and are related to the field invariants in the following way:
\begin{equation}
a^2 + b^2 = \frac{1}{2} F_{\mu \nu} F^{\mu \nu}  \quad ,  \quad 
a b = \frac{1}{2} F_{\mu \nu} \tilde{F}^{\mu \nu} \, .
\end{equation}
The leading term in the derivative expansion is obtained by  simply replacing the
the constants $a$ and $b$ by the corresponding space dependent quantities inside these integral expressions, and then obtaining the effective action from the effective Lagrangian by integrating over spacetime. For the class of backgrounds  that we are considering, the corresponding substitution is
\begin{eqnarray}
a &\longrightarrow& a(r) = 2 g(r)  \, , \nn \\
b &\longrightarrow& b(r) = 2 g(r) + r g'(r)  \, ,
\end{eqnarray}
such that,
\begin{equation}
\Gamma_{\rm DE} = 2 \pi^2 \int_0^\infty r^3\,dr \, \mathcal{L} (a(r), b(r)) \, .
\end{equation}
At first sight, one would expect the derivative expansion to be a good approximation at large mass, similar to the large mass expansion, since the expansion over derivatives in (\ref{der}) is balanced by inverse powers of $m$. However, the situation is more subtle than that naive expectation, as the derivative expansion expression at a given order is a resummation of all powers of the field strength with a fixed number of derivatives. Thus, one might expect that the derivative expansion is better than the large mass expansion for smaller values of the mass. Indeed, in previous work on the partial-wave cutoff method in scalar theories it was found that even
 the leading term of the derivative expansion  provides a surprisingly accurate approximation to the mass dependence of  effective action~\cite{Dunne:2007mt}, even for very small mass. In the next section we study this question for spinor theories and find an interesting difference.

\section{Approximation methods versus exact calculation}

In this section we use exact results obtained from our partial-wave cutoff method to probe the validity of the large-mass and deriative expansion approximations, for spinor QED.
In order to emphasize 
the similarities and differences between the scalar and spinor cases, we also show some results for the 
scalar action as well.

\subsection{The scalar case}

To exemplify how the partial-wave cutoff method works in the scalar theory,  we use the background field  given by $g_1(r)$, choosing different values of the range parameter $\xi$. We present
a comparison between the the exact effective
action as calculated with the partial-wave cutoff  method, the large-mass expansion and the derivative
expansion.
\begin{figure}[htb]
\begin{center}
\includegraphics[width=0.6\textwidth]{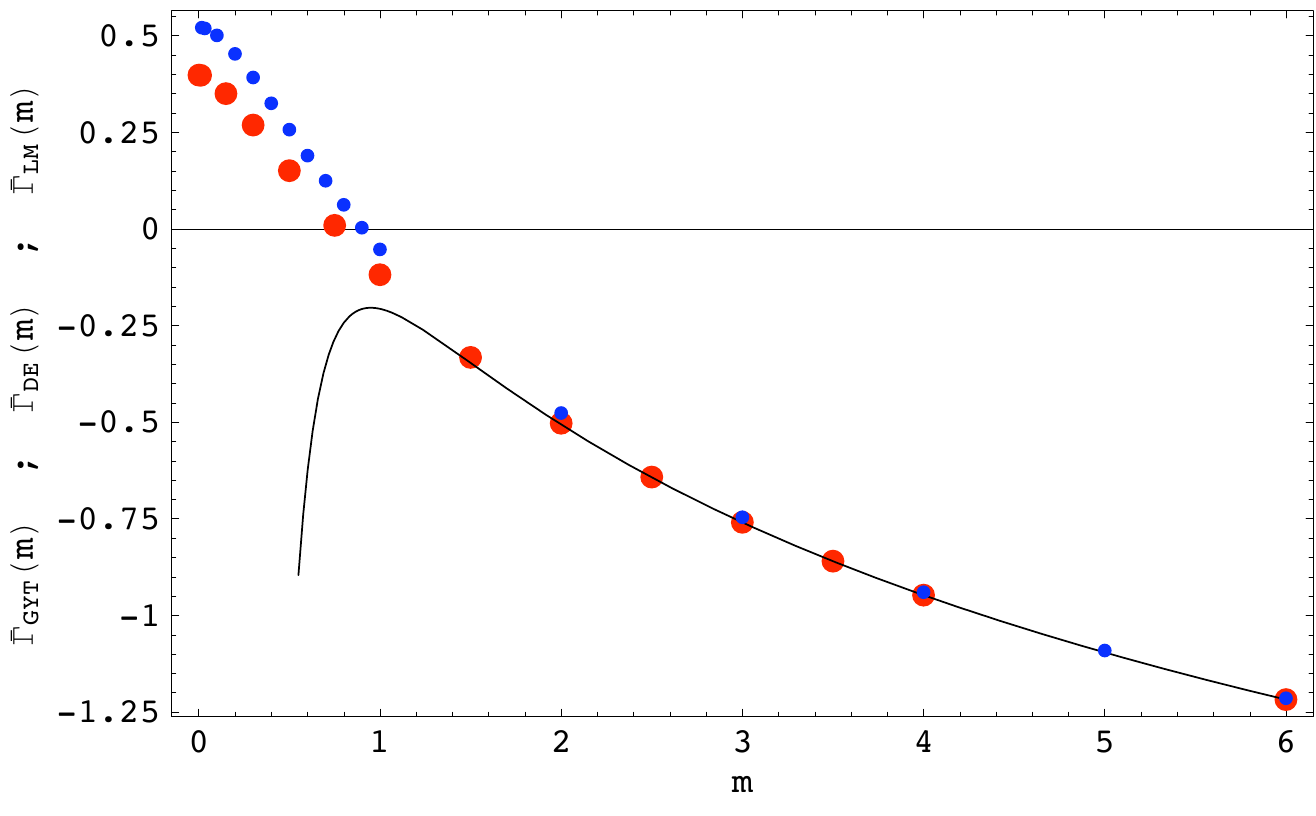}
\includegraphics[width=0.6\textwidth]{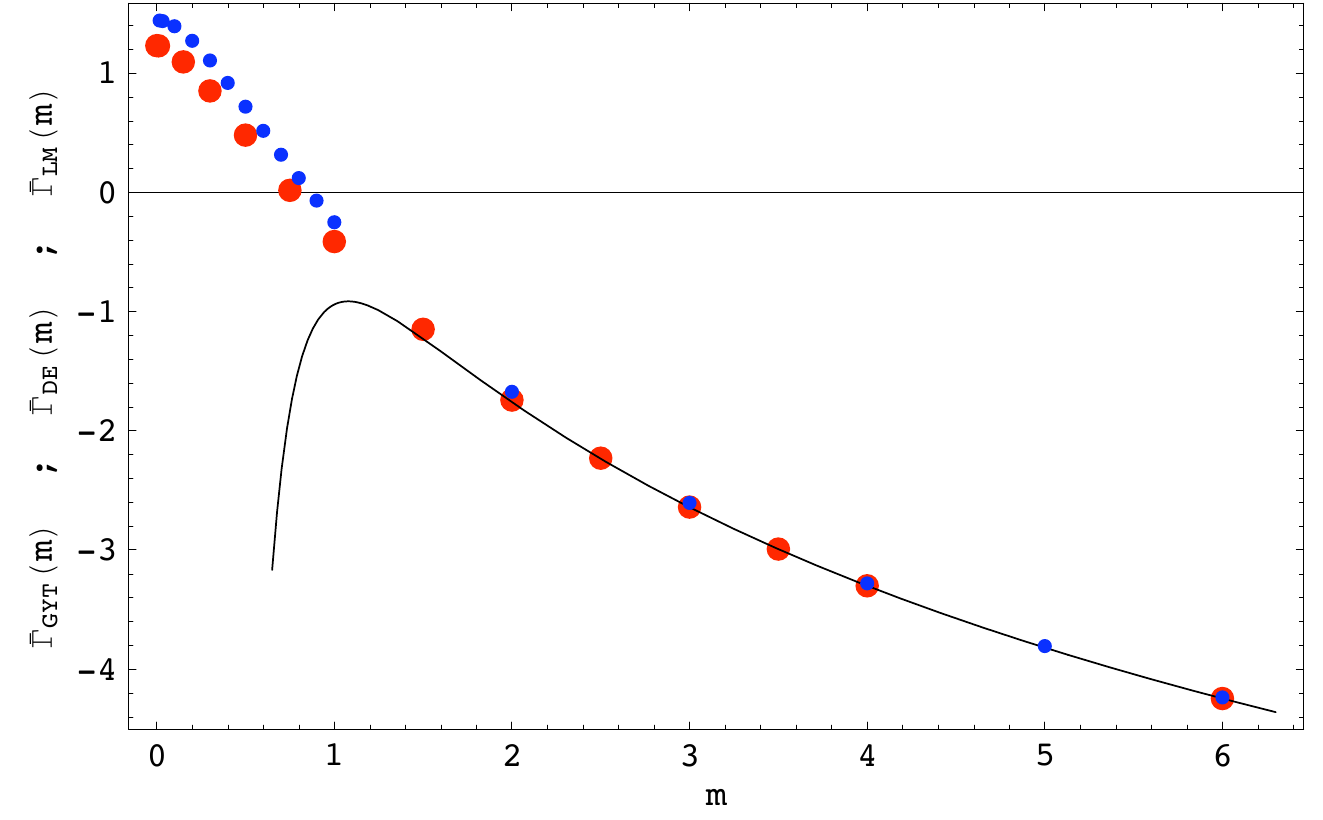}
\includegraphics[width=0.6\textwidth]{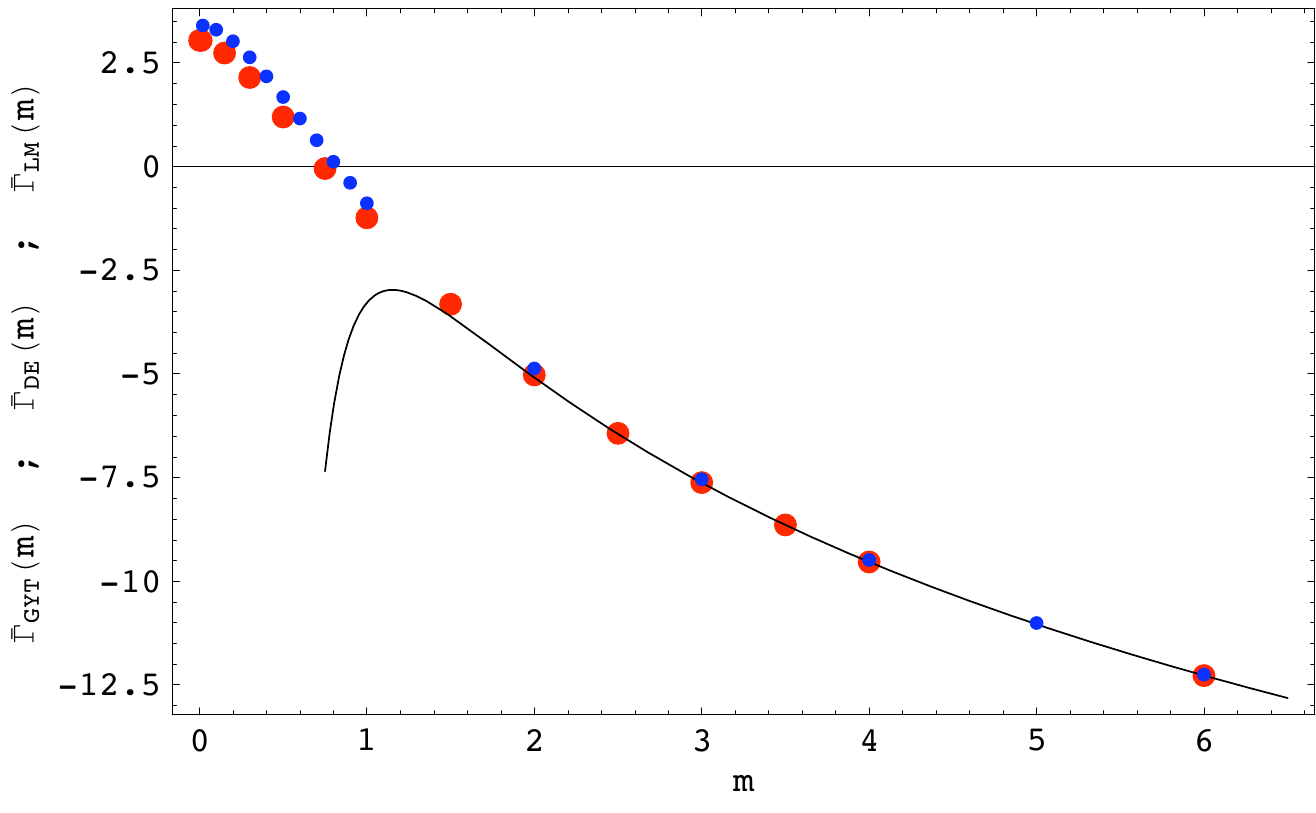}
\end{center}
\caption{  The graphs show,  for the case of scalar QED, our exact effective action, $\tilde{\Gamma}_{{\rm GYT}}(m)$ (big-dots/red),  the derivative expansion expression $\tilde{\Gamma}_{{\rm DE}}(m)$ (small-dots/blue),  and  the large mass expansion expression, $\tilde{\Gamma}_{{\rm LM}}(m)$ (solid-line).  
 We use the profile function $g_1(r)= B (1 - {\rm
{Tanh}}( \beta (\sqrt{B} r  - \xi ) ))$ with  $B = 1$, $\beta = 1$. We show results for three values of the \emph{range} parameter: $\xi = 1$ (upper graph), $\xi = 3/2$ (middle graph) and  $\xi = 2$ (lower graph). Notice that the derivative expansion is a reasonable approximation even at zero mass. } 
\label{fig3}
\end{figure}
From the plots in  Figure \ref{fig3},  we make the following observations:
\begin{itemize}
\item The large-mass expansion result agrees very well with the exact result,  already for $m \sim 2$. 
\item The leading order derivative expansion is good at large $m$, and also provides a very good approximation
to the effective action for large values of the parameter  $\xi$.
\item In particular, as long as the steepness parameter $\xi$ is large enough, the derivative expansion provides
accurate results in both the large-mass and small-mass regimes.
\end{itemize}

\subsection{The spinor case}

\begin{figure}[htb]
\begin{center}
\includegraphics[width=0.8\textwidth]{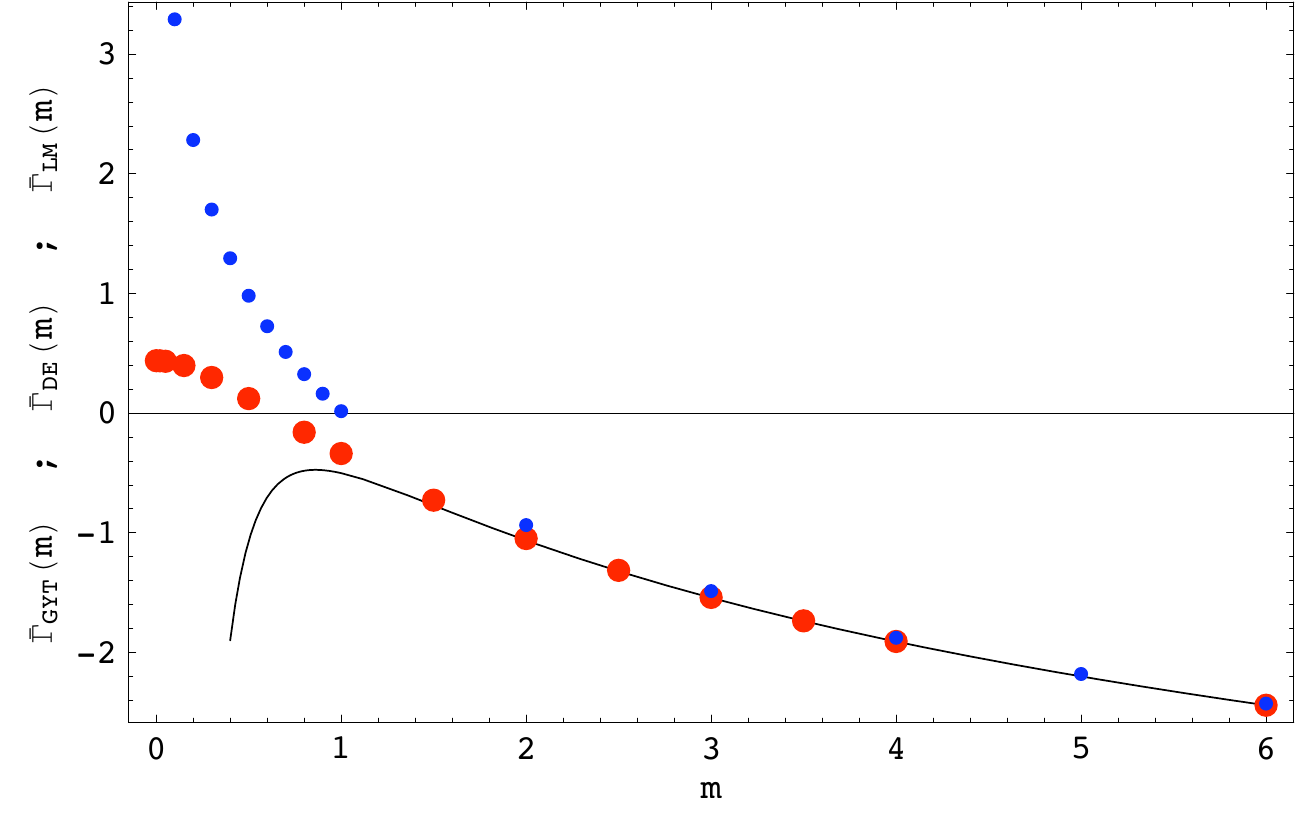}
\end{center}
\caption{For spinor QED, this plot shows the exact effective action, $\tilde{\Gamma}_{{\rm GYT}}(m)$ (big-dots/red), the derivative expansion expression, $\tilde{\Gamma}_{{\rm DE}}(m)$ (small-dots/blue),    and   the large mass expansion expression, $\tilde{\Gamma}_{{\rm LM}}(m)$ (solid-line).  We use the same profile function as in the upper plot of Figure \ref{fig3} [which is for scalar QED]: $g_1(r)= B (1 - {\rm
{Tanh}}( \beta (\sqrt{B} r  - \xi ) ))$ with  $B = 1$, $\beta = 1$ and $\xi = 1$.  Note the very different behavior at small mass compared to the scalar QED case.}
 \label{fig4}
\end{figure}

In this section we compare and analyse the different approximation methods against the GYT method.  The first background configuration we investigate is $g_1(r)$, the plots
shown in fig.~(\ref{fig4}) correspond to the GYT method, the large-mass expansion and the derivative expansion.
From the plot in  Figure \ref{fig4},  we make the following observations:
\begin{itemize}
\item The partial-wave cutoff method produces a finite value for the effective action in the small-mass regime.
\item As in scalar QED, for spinor QED the large-mass expansion result agrees very well with the exact result,  already for $m \sim 2$. 
\item As in scalar QED, for spinor QED the leading order derivative expansion is good at large $m$. However, the leading order  derivative expansion result diverges in the small-mass regime.
\end{itemize}
To understand why the leading-order derivative expansion fails in the small-mass limit, we analyse equation
(\ref{euclidean-EHsp}) in the small-mass regime. The small-mass regime is obtained by taking $s \to \infty$, which gives:
\begin{eqnarray}
\mathcal{L}_{{\rm spinor}}( a, b)
&\sim&  -\frac{1}{8 \pi^2} \Big[ a b  - \frac{ 1}{3}\,( a^2 + b^2 ) \Big] \int_0^{\infty} \, \frac{ds}{s}\, e^{-m^2 s }
 \nonumber \\
 &\sim&  \frac{1}{4 \pi^2} \Big[ a b  - \frac{ 1}{3}\,( a^2 + b^2 ) \Big] \ln m  \quad .
 \label{spinorde}
\end{eqnarray}
We recognize one $\ln m$ divergence proportional to $F_{\mu\nu}F_{\mu\nu}$, and another proportional to  $F_{\mu\nu}\tilde{F}_{\mu\nu}$. Contrast this with scalar QED where 
\begin{eqnarray}
\mathcal{L}_{{\rm scalar}}( a, b)
&\sim&  \frac{1}{16 \pi^2} \Big[ 0  + \frac{ 1}{6}\,( a^2 + b^2 ) \Big] \int_0^{\infty} \, \frac{ds}{s}\, e^{-m^2 s }
 \nonumber \, , \\
 &\sim& - \frac{1}{8 \pi^2} \Big[  \frac{ 1}{6}\,( a^2 + b^2 ) \Big] \ln m   \quad .
\end{eqnarray}
which has a $\ln m$ divergence proportional to $F_{\mu\nu}F_{\mu\nu}$, but none proportional to  $F_{\mu\nu}\tilde{F}_{\mu\nu}$.  This is simply a reflection of the fact that zero modes can occur in spinor QED but not in scalar QED, with the $F_{\mu\nu}\tilde{F}_{\mu\nu}$ term determining the number of zero modes.

But note that  $\mathcal{L}_{{\rm spinor}}( a, b) = \mathcal{L}_{{\rm spinor}}( -a, -b)$. Therefore, (\ref{spinorde}) really should read
\begin{eqnarray}
\mathcal{L}_{{\rm spinor}}( a, b)
 \sim  \frac{1}{4 \pi^2} \Big[ |a b|  - \frac{ 1}{3}\,( a^2 + b^2 ) \Big] \ln m  \quad .
 \label{spinorde2}
\end{eqnarray}
Therefore, our naive application of the leading-order derivative expansion includes a term in the effective action that behaves in the small mass limit as 
\begin{eqnarray}
\left(\int d^4 x\,  |a(r)b(r)|\right)\ln m
\label{1}
\end{eqnarray}
instead of the correct form
\begin{eqnarray}
\left(\int d^4 x\,  a(r)b(r)\right)\ln m
\label{2}
\end{eqnarray}
It is this latter form that counts the number of zero modes, and appears in the counter-term. To show the effect of this, consider the radial profile function $g_1(r)$ for which $\int d^4 x\, F_{\mu\nu}\tilde{F}_{\mu\nu}=0$, indicating the absence of zero modes. For this profile function, this integral vanishes because the integrand changes sign. But the derivative expansion expression does not allow for such changes of sign, and so we effectively compute $\int d^4 x\, \left | F_{\mu\nu}\tilde{F}_{\mu\nu}\right |$, which is non-zero. 

\begin{figure}[htb]
\begin{center}
\includegraphics[width=0.6\textwidth]{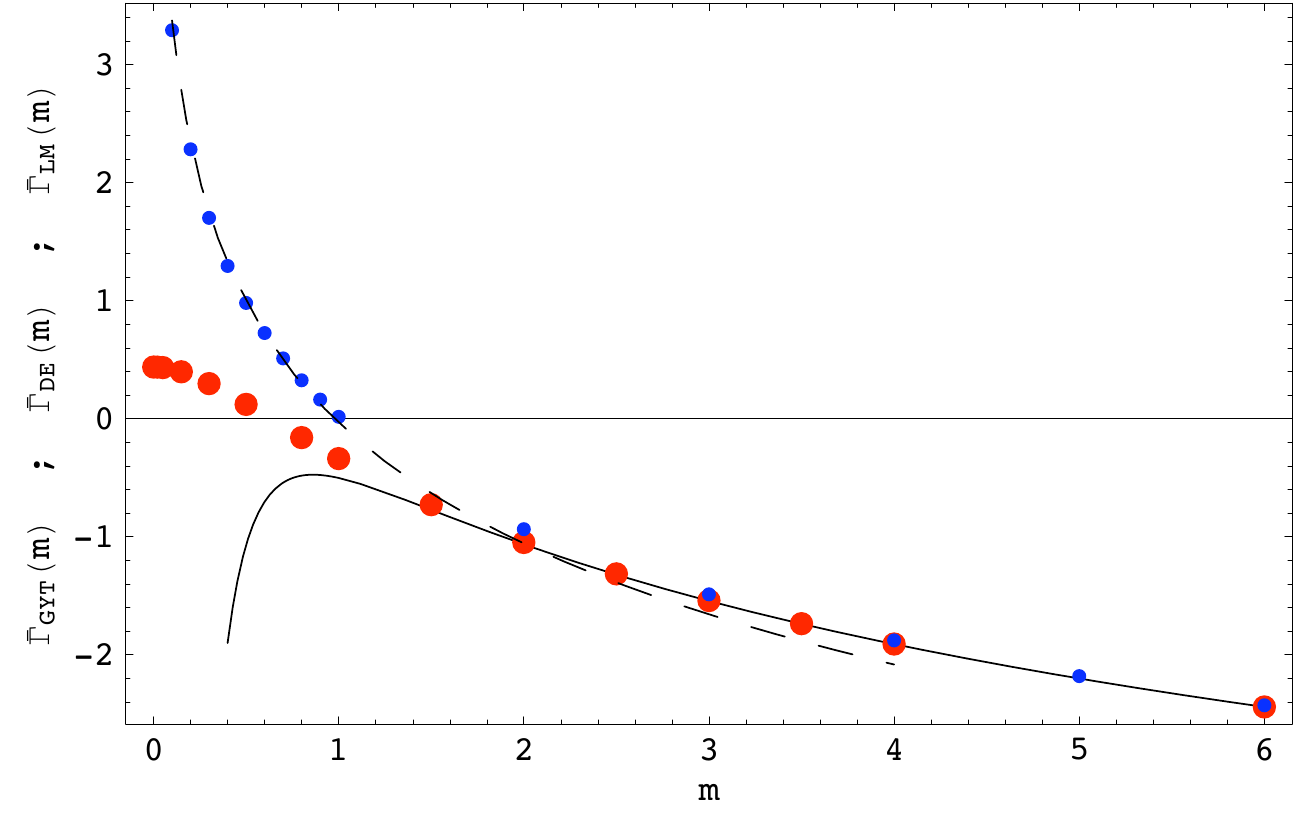}
\includegraphics[width=0.6\textwidth]{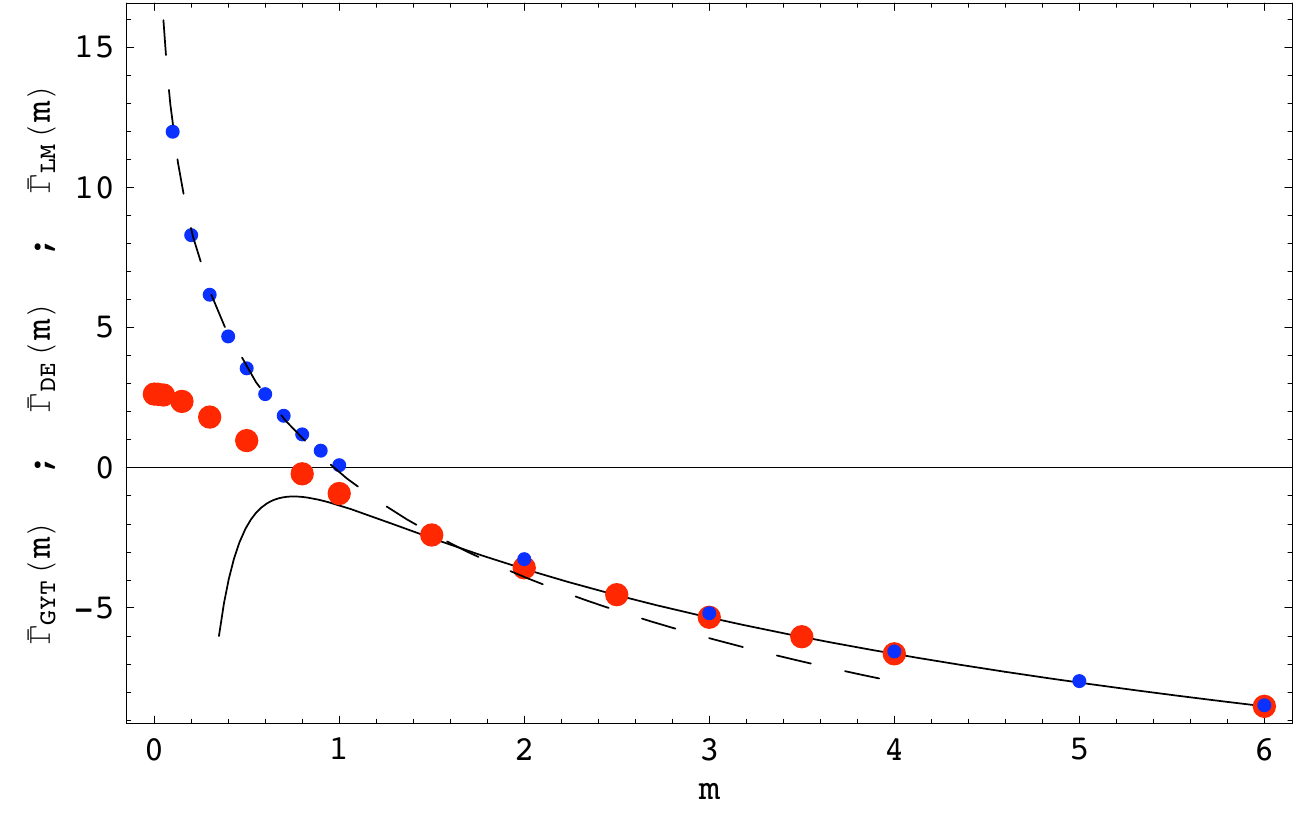}
\includegraphics[width=0.6\textwidth]{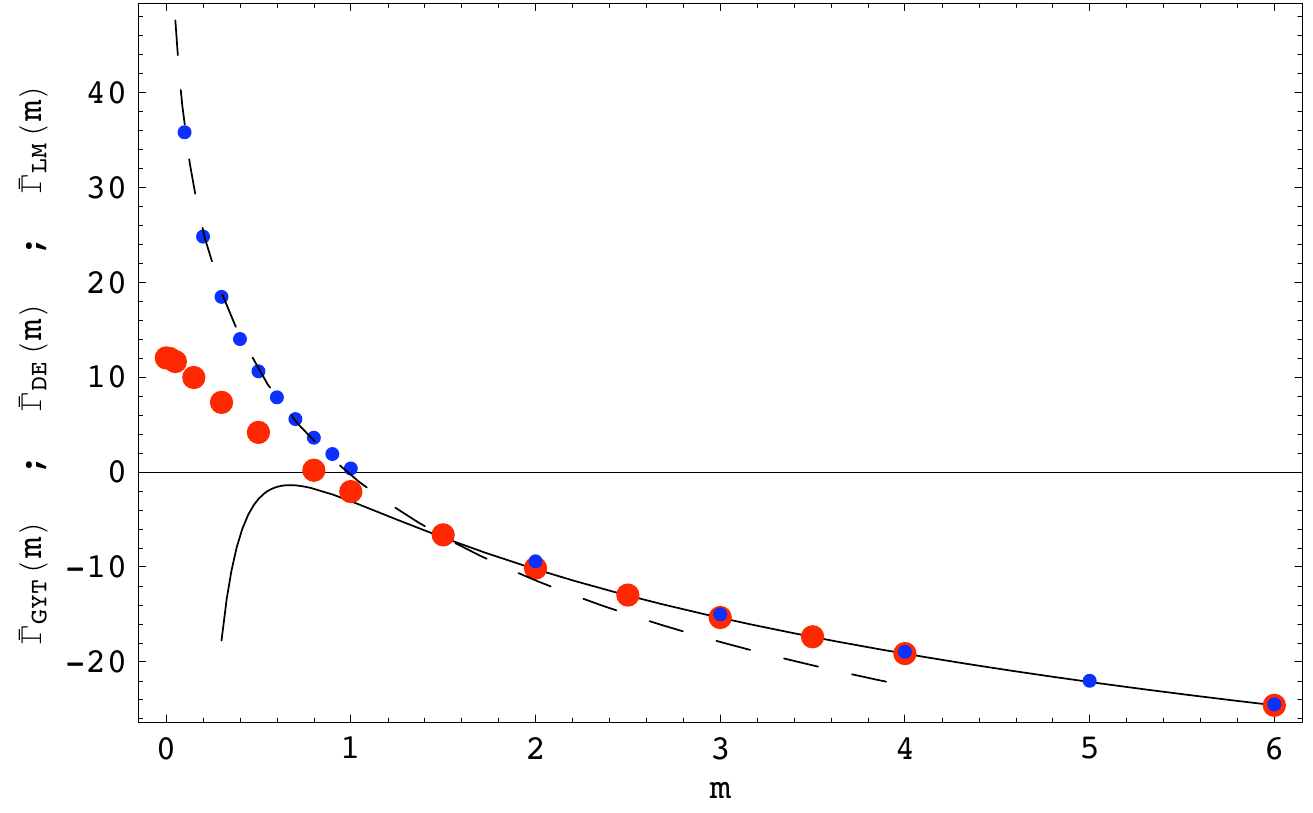}
\end{center}
\caption{For spinor QED, these graphs show  the exact effective action, $\tilde{\Gamma}_{{\rm GYT}}(m)$ (big-dots/red), the derivative expansion expression, $\tilde{\Gamma}_{{\rm DE}}(m)$ (small-dots/blue),     and   the large-mass expansion expression $\tilde{\Gamma}_{{\rm LM}}(m)$ (solid-line). The dashed line represents the residual logarithm $f(m) = \frac{1}{4 \pi^2} (\int d^4 x |a(r)b(r)|) \ln m$.   We use the profile function $g_1(r)= B (1 - {\rm
{Tanh}}( \beta (\sqrt{B} r  - \xi ) ))$  with  $B = 1$, $\beta = 1$. We show results for three values of the \emph{range} parameter: $\xi = 1$ (upper graph), $\xi = 3/2$ (middle graph) and  $\xi = 2$ (lower graph). Note that the divergent behavior at small mass is well fitted by the residual logarithm, as described in the text.} 
\label{fig5}
\end{figure}

This mis-match is demonstrated clearly  in Figures \ref{fig5},  which show that the divergent small-mass behavior of the derivative expansion corresponds exactly to
\begin{equation}
\tilde{\Gamma}_{{\rm DE}}^{{\rm spinor}}(m) \sim f(m) = \Big( \frac{1}{4 \pi^2} \int d^4 x \;  |a(r)b(r)| \Big) \, \ln m \quad ; \quad m \to 0 \, .
\end{equation}
The number of zero-modes is given by $\frac{1}{4 \pi^2} \int d^4 x  a(r) b(r)  $ and it is always equal to zero for  the
 backgrounds of the class $g_1(r)$, however, $\frac{1}{4 \pi^2} \int d^4 x |a(r)b(r)|$ does not vanish and  this adds an
incorrect residual logarithmic dependence to the derivative expansion, as we have shown.
No such mis-match occurs for scalar QED because the small $m$ behavior of the derivative expansion expression only involves the combination $F_{\mu\nu}F_{\mu\nu}\propto [a^2(r)+b^2(r)]$, which is always positive. Thus there is no $\ln m$ divergence in the small $m$ behavior of the derivative expansion plots shown in Figure \ref{fig3}
for scalar QED.

\begin{figure}[tb]
\begin{center}
\includegraphics[width=0.6\textwidth]{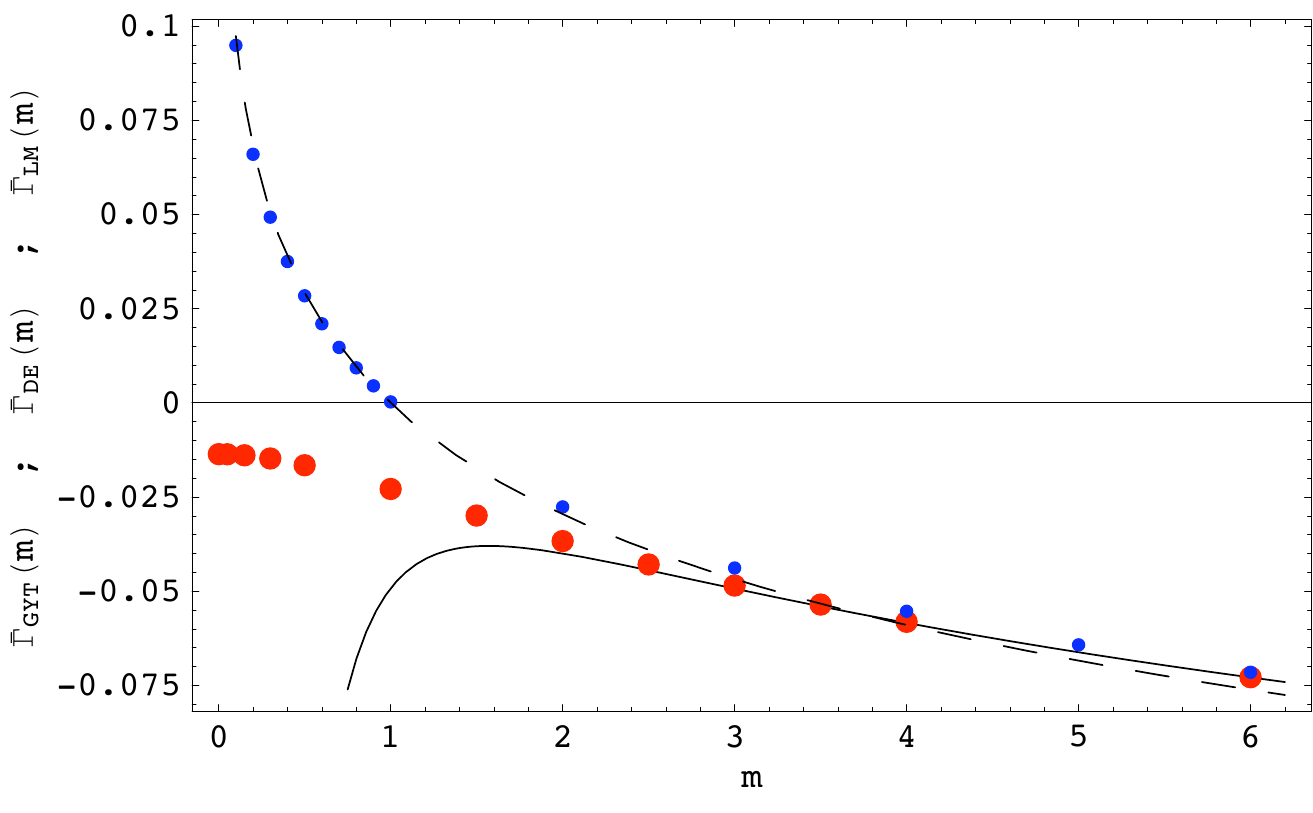}
\includegraphics[width=0.6\textwidth]{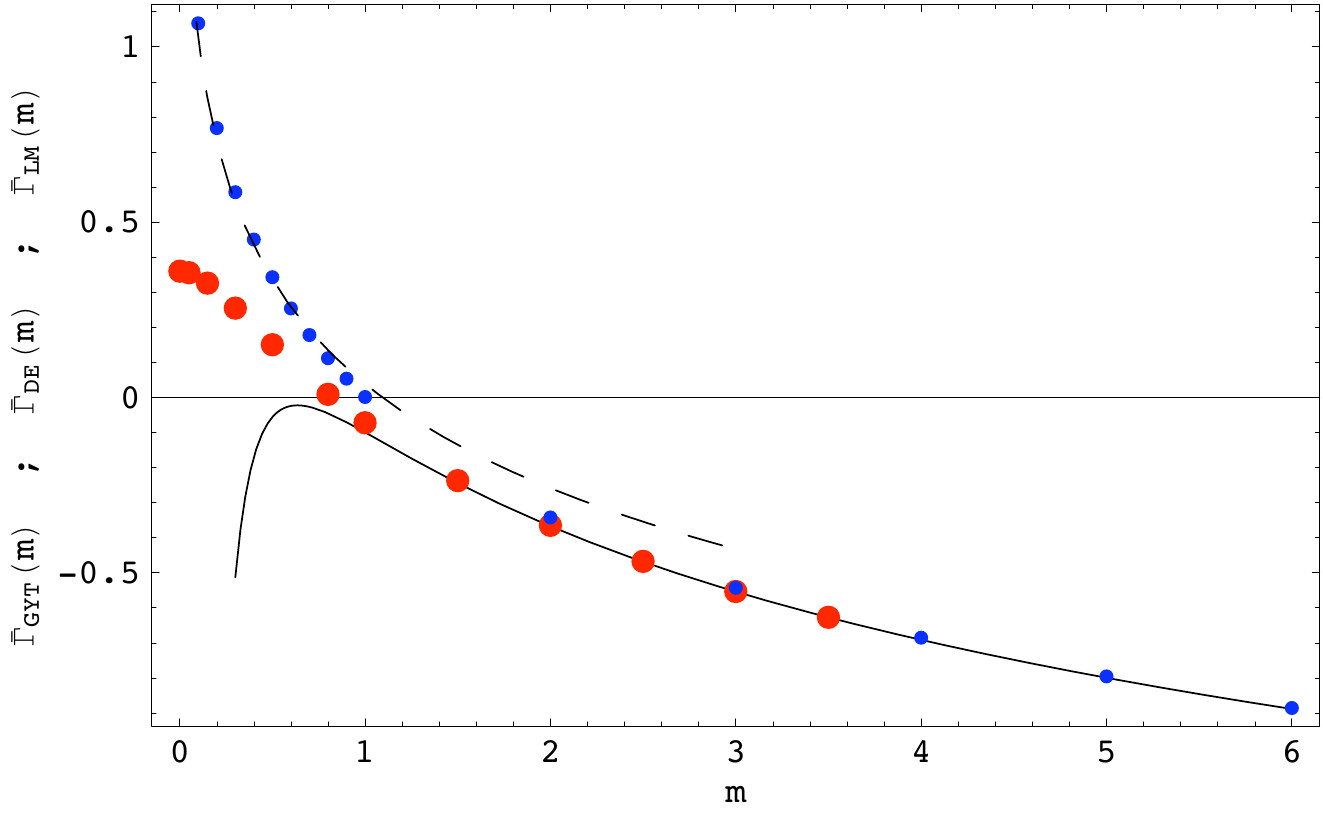}
\includegraphics[width=0.6\textwidth]{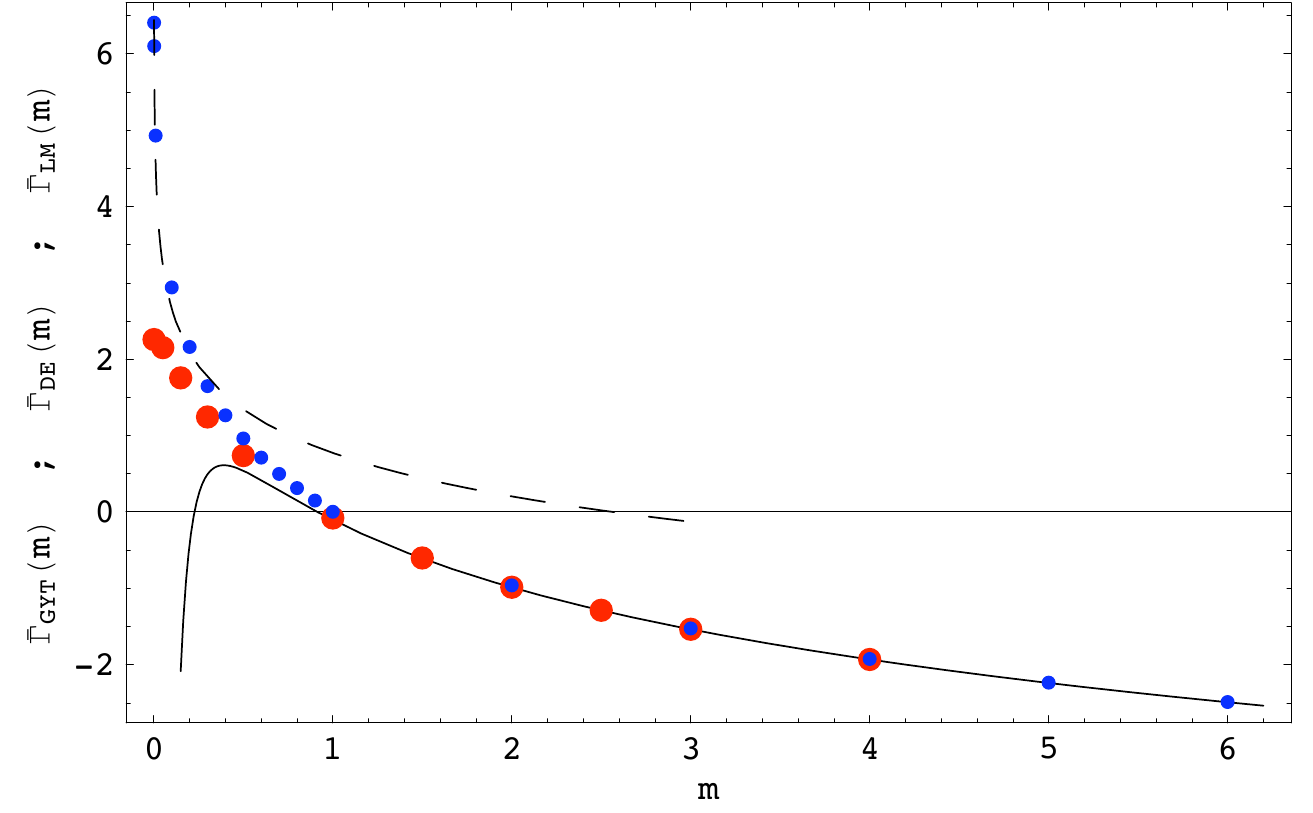}
\end{center}
\caption{  For spinor QED, these graphs show  the exact effective action, $\tilde{\Gamma}_{{\rm GYT}}(m)$ (big-dots/red), the derivative expansion expression, $\tilde{\Gamma}_{{\rm DE}}(m)$ (small-dots/blue),     and   the large-mass expansion expression $\tilde{\Gamma}_{{\rm LM}}(m)$ (solid-line). The dashed line corresponds to $\frac{1}{4 \pi^2} (\int d^4 x | a(r) b(r)|) \ln m$.  
 We use the profile function $g_2 (r) = \nu  e^{-\alpha r^2}/(\rho^2 + r^2)$ with   $\nu = 1$ and $\rho = 1$. We show three values
 of the decay rate parameter:  $\alpha = 1$ (upper graph), $\alpha = 1/20$ (middle graph) and $\alpha = 1/400$ (lower graph). Note that the divergent behavior at small mass is well fitted by the residual logarithm, as described in the text.} 
 \label{fig6}
\end{figure}

The second background we examined is the one given by the profile function $g_2(r)$.
Setting different values of the parameter $\alpha$, we can control the rate of decay
of the potential.  In figure~\ref{fig6},  we present a comparison between the different calculation
methods for this kind of background. 
We  corroborate once more how the derivative expansion fails at the zero-mass limit. Also, the derivative expansion  shows better accuracy in a wider mass-range as the
for those fields with slower variation (small $\alpha$), as expected.
It also evident that the residual logarithm is not the dominant term  in the derivative expansion when we move from the small-mass  regime in to the large-mass regime. 

\section{Conclusion}

 We have  extended  the partial-wave
cutoff method to spinor theories with nontrivial [and non-self-dual] radially symmetric gauge backgrounds. Different background fields were tested, resulting always in accurate values of the effective action for both
the large-mass and small-mass regimes.  We provided an example of how our  method  allows one to systematically 
investigate how the effective action responds to different characteristics of the background fields, such as range, amplitude or rate of variation. 
We have also analyzed how certain approximation methods compare with these exact results, in
the different mass regimes, also comparing the scalar case with the the spinor case. Specifically, we have tested the large-mass expansion and the derivative expansion. We have shown that the large mass expansion works extremely well, as in the scalar case. However, the derivative expansion behaves in a different manner in the small-mass regime for the spinor theory. We have explained this fact, both qualitatively and quantitatively, as resulting from the changing sign of the local quantity $F_{\mu\nu}\tilde{F}_{\mu\nu}$, and which in the derivative expansion approximation is assumed to be constant and therefore of fixed sign. We have also shown that this is directly related to the appearance of fermion zero modes.
 
\bigskip
GD and AH were supported by the US DOE grant DE-FG02-92ER40716,  and HM was supported by the National Research
Foundation of Korea (NRF) funded by the Ministry of Education, Science and Technology (No 2010-0011223).



\begin{thebibliography}{00}  

\bibitem{Jackiw:1974cv}
  R.~Jackiw,
  ``Functional evaluation of the effective potential,''
  Phys.\ Rev.\  {\bf D9}, 1686 (1974).
  
\bibitem{Iliopoulos:1974ur}
  J.~Iliopoulos, C.~Itzykson, A.~Martin,
  ``Functional Methods and Perturbation Theory,''
  Rev.\ Mod.\ Phys.\  {\bf 47}, 165 (1975).

\bibitem{Heisenberg:1935qt}
  W.~Heisenberg and H.~Euler,
  ``Consequences of Dirac's theory of positrons,''
  Z.\ Phys.\  {\bf 98}, 714 (1936).

\bibitem{Schwinger:1951nm}
  J.~S.~Schwinger,
  ``On gauge invariance and vacuum polarization,''
  Phys.\ Rev.\  {\bf 82}, 664 (1951).
  
\bibitem{Dunne:2004nc}
  G.~V.~Dunne,
  ``Heisenberg-Euler effective Lagrangians: Basics and extensions,''
  in Ian Kogan Memorial Collection, {\it From Fields to Strings: Circumnavigating Theoretical Physics'}, 
M. Shifman et al (ed.), vol. {\bf 1},  445-522
  [\hhref{hep-th/0406216}].


\bibitem{Novikov:1983gd}
  V.~A.~Novikov, M.~A.~Shifman, A.~I.~Vainshtein, V.~I.~Zakharov,
  ``Calculations in External Fields in Quantum Chromodynamics. Technical Review,''
  Fortsch.\ Phys.\  {\bf 32}, 585 (1984).

\bibitem{Dunne:2004sx}
  G.~V.~Dunne, J.~Hur, C.~Lee and H.~Min,
  ``Precise quark mass dependence of instanton determinant,''
  Phys.\ Rev.\ Lett.\  {\bf 94}, 072001 (2005)
  [\hhref{hep-th/0410190}].

\bibitem{Dunne:2005te}
  G.~V.~Dunne, J.~Hur, C.~Lee and H.~Min,
  ``Calculation of QCD instanton determinant with arbitrary mass,''
  Phys.\ Rev.\  D {\bf 71}, 085019 (2005)
  [\hhref{hep-th/0502087}].


\bibitem{Dunne:2006ac}
  G.~V.~Dunne, J.~Hur and C.~Lee,
  ``Renormalized effective actions in radially symmetric backgrounds. I:
  Partial wave cutoff method,''
  Phys.\ Rev.\  D {\bf 74}, 085025 (2006)
  [\hhref{hep-th/0609118}].

\bibitem{Dunne:2007mt}
  G.~V.~Dunne, J.~Hur, C.~Lee and H.~Min,
 ``Renormalized Effective Actions in Radially Symmetric Backgrounds: Exact
  Calculations Versus Approximation Methods,''
  Phys.\ Rev.\  D {\bf 77}, 045004 (2008)
  [\hhref{0711.4877}].
  
\bibitem{Hur:2008yg}
  J.~Hur and H.~Min,
  ``A Fast Way to Compute Functional Determinants of Radially Symmetric Partial
  Differential Operators in General Dimensions,''
  Phys.\ Rev.\  D {\bf 77}, 125033 (2008)
  [\hhref{0805.0079}].


\bibitem{'tHooft:1976fv}
  G.~'t Hooft,
  ``Computation of the quantum effects due to a four-dimensional
  pseudoparticle,''
  Phys.\ Rev.\  D {\bf 14}, 3432 (1976)
  [Erratum-ibid.\  D {\bf 18}, 2199 (1978)].


\bibitem{Jackiw:1977pu}
  R.~Jackiw and C.~Rebbi,
  ``Spinor analysis of Yang-Mills theory,''
  Phys.\ Rev.\  D {\bf 16}, 1052 (1977).
  
\bibitem{Brown:1977bj}
  L.~S.~Brown, R.~D.~Carlitz and C.~Lee,
  ``Massless Excitations In Instanton Fields,''
  Phys.\ Rev.\  D {\bf 16}, 417 (1977).


\bibitem{Carlitz:1978xu}
  R.~D.~Carlitz, C.~Lee,
  ``Physical Processes In Pseudoparticle Fields: The Role Of Fermionic Zero Modes,''
  Phys.\ Rev.\  {\bf D17}, 3238 (1978).

\bibitem{Gelfand:1959nq}
  I.~M.~Gelfand and A.~M.~Yaglom,
  ``Integration in functional spaces and it applications in quantum physics,''
  J.\ Math.\ Phys.\  {\bf 1}, 48 (1960).


\bibitem{Hur:2010bd}
  J.~Hur, C.~Lee and H.~Min,
  ``Some chirality-related properties of the 4-D massive Dirac propagator and
  determinant in an arbitrary gauge field,''
  Phys.\ Rev.\  {\bf D82}, 085002 (2010)
[\hhref{1007.4616}].



\bibitem{Adler:1972qq}
  S.~L.~Adler,
  ``Massless, Euclidean quantum electrodynamics on the five-dimensional unit
  hypersphere,''
  Phys.\ Rev.\  D {\bf 6}, 3445 (1972)
  [Erratum-ibid.\  D {\bf 7}, 3821 (1973)].


\bibitem{Adler:1974nd}
  S.~L.~Adler,
  ``Massless electrodynamics in the one photon mode approximation'',
  Phys.\ Rev.\  D {\bf 10}, 2399 (1974)
  [Erratum-ibid.\  D {\bf 15}, 1803 (1977)].


\bibitem{Bogomolny:1981qv}
  E.~B.~Bogomolny and Yu.~A.~Kubyshin,
  ``Asymptotical Estimates For Graphs With A Fixed Number Of Fermionic Loops In
  Quantum Electrodynamics. The Choice Of The Form Of The Steepest Descent
  Solutions,''
  Sov.\ J.\ Nucl.\ Phys.\  {\bf 34}, 853 (1981)
  [Yad.\ Fiz.\  {\bf 34}, 1535 (1981)].


\bibitem{Bogomolny:1982ea}
  E.~B.~Bogomolny and Yu.~A.~Kubyshin,
  ``Asymptotical Estimates For Graphs With A Fixed Number Of Fermionic Loops In
  Quantum Electrodynamics. The Extremal Configurations With The Symmetry Group
  O(2) X O(3),''
  Sov.\ J.\ Nucl.\ Phys.\  {\bf 35}, 114 (1982)
  [Yad.\ Fiz.\  {\bf 35}, 202 (1982)].




\bibitem{Fry:2003uy}
  M.~P.~Fry,
  ``Fermion determinant for general background gauge fields,''
  Phys.\ Rev.\  D {\bf 67}, 065017 (2003)
  [\hhref{hep-th/0301097}].



\bibitem{Fry:2010cd}
  M.~P.~Fry,
  ``Nonperturbative results for the mass dependence of the QED fermion
  determinant,''
  Phys.\ Rev.\  D {\bf 81}, 107701 (2010)
  [\hhref{1005.4849}].




\bibitem{Cangemi:1994by}
  D.~Cangemi, E.~D'Hoker, G.~V.~Dunne,
  ``Derivative expansion of the effective action and vacuum instability for QED in (2+1)-dimensions,''
  Phys.\ Rev.\  {\bf D51}, 2513-2516 (1995)
  [\hhref{hep-th/9409113}].



\bibitem{Gusynin:1995bc}
  V.~P.~Gusynin, I.~A.~Shovkovy,
  ``Derivative expansion for the one loop effective Lagrangian in QED,''
  Can.\ J.\ Phys.\  {\bf 74}, 282-289 (1996)
  [\hhref{hep-ph/9509383}].


\bibitem{Salcedo:2000hp}
  L.~L.~Salcedo,
  ``Derivative expansion for the effective action of chiral gauge fermions: The Normal parity component,''
  Eur.\ Phys.\ J.\  {\bf C20}, 147 (2001)
  [\hhref{hep-th/0012166}];
  L.~L.~Salcedo,
  ``Derivative expansion for the effective action of chiral gauge fermions. the abnormal parity component,''
  Eur.\ Phys.\ J.\  {\bf C20}, 161 (2001)
  [\hhref{hep-th/0012174}].










\end{thebibliography}
\end{document}